# RECOGNIZING DIFFERENT TYPES OF STOCHASTIC PROCESSES

JONG U. KIM AND LASZLO B. KISH

*Department of Electrical and Computer Engineering, Texas A&M University, College Station, TX 77843-3128, USA*

We propose a new cross-correlation method that can recognize independent realizations of the same type of stochastic processes and can be used as a new kind of pattern recognition tool in biometrics, sensing, forensic, security and image processing applications. The method, which we call *bispectrum correlation coefficient* method, makes use of the cross-correlation of the bispectra. Three kinds of cross-correlation coefficients are introduced. To demonstrate the new method, six different random telegraph signals are tested, where four of them have the same power density spectrum. It is shown that the three coefficients can map the different stochastic processes to specific sub-volumes in a cube.

## 1. Introduction

Identification and pattern recognition techniques are of crucial importance in biometric, sensing, forensic, security and image processing applications. In this Letter, we will introduce a new method, which is a *process recognition tool*, recognizing different types of stochastic processes. The method is based on the bispectra (a higher order statistical tool) which have recently been applied to identity gases by fluctuation-enhanced gas sensing [1, 2].

We name this new process recognition tool *bispectrum correlation coefficient* (BCC) method because it utilizes normalized cross-correlation coefficients based on the bispectra of the process realizations.

Conventional cross-correlation techniques recognize only the *same realization* of a stochastic process, and they give zero value for the *independent realizations of the same process* or for two different processes. Consequently, cross-correlation techniques cannot distinguish between the case of *two independent realizations of the same process* and that of *two different processes*. In this letter, we will show that the BCC method is useful for the identification of stochastic processes even though their power density spectra (PDS) or amplitude distribution functions are indistinguishable.

## 2. Bispectrum correlation coefficient

The bispectrum for a stationary signal $x(k)$ is defined as [3]:

$$S_3(\omega_1,\omega_2) = \sum_{\tau_1=-\infty}^{\infty} \sum_{\tau_2=-\infty}^{\infty} \{E[x(k)x(k+\tau_1)x(k+\tau_2)] \cdot \exp[-j(\omega_1\tau_1 + \omega_2\tau_2)]\} \quad, \tag{1}$$

where $\tau$ and $\omega$ are discrete time and the angular frequency, respectively, and $E[...]$ means ensemble average. The signal is supposed to be stationary with zero mean. Due to the symmetry of the bispectrum [3], the whole information lies in the non-redundant region:

$$\omega_1 \leq \omega_2, \quad 0 \leq \omega_2, \quad \text{and} \quad (\tau_{i+1} - \tau_i)(\omega_1 + \omega_2) \leq \pi \quad. \tag{2}$$

To calculate the bispectrum from a stationary signal of finite length, we can use the so-called *direct conventional method* [3,4]:

$$S_3(\omega_1,\omega_2) = \frac{X(\omega_1)X(\omega_2)X^*(\omega_1+\omega_2)}{N} \quad, \tag{3}$$

where $N$ is the number of samples, the asterisk represents complex conjugate and $X(\omega)$ is the Fourier transform:

$$X(\omega) = \sum_{k=0}^{M-1} x(k)\exp(-jk\omega) \quad, \tag{4}$$

where $M$ is the length of a sample. As implied in Eq. (3), the bispectrum for a real-valued signal is a two-dimensional matrix with complex number elements.

To study the correlations between two bispectra, we introduce the BCC which has three different types. The "*real*" bispectrum correlation coefficient (RBCC) is:

$$\text{RBCC} = \frac{\sum \text{Re}[S_3^i(\omega_1,\omega_2)] \cdot \text{Re}[S_3^j(\omega_1,\omega_2)]}{\sqrt{\sum \{\text{Re}[S_3^i(\omega_1,\omega_2)]\}^2 \cdot \sum \{\text{Re}[S_3^j(\omega_1,\omega_2)]\}^2}} \quad, \tag{5a}$$

the "*imaginary*" bispectrum correlation coefficient (IBCC) is:

$$\text{IBCC} = \frac{\sum \text{Im}[S_3^i(\omega_1,\omega_2)] \cdot \text{Im}[S_3^j(\omega_1,\omega_2)]}{\sqrt{\sum \{\text{Im}[S_3^i(\omega_1,\omega_2)]\}^2 \cdot \sum \{\text{Im}[S_3^j(\omega_1,\omega_2)]\}^2}} \quad, \tag{5b}$$

and the "*magnitude*" bispectrum correlation coefficient (MBCC) is:

$$\text{MBCC} = \frac{\sum |S_3^i(\omega_1,\omega_2)| \cdot |S_3^j(\omega_1,\omega_2)|}{\sqrt{\sum |S_3^i(\omega_1,\omega_2)|^2 \cdot \sum |S_3^j(\omega_1,\omega_2)|^2}} \quad, \tag{5c}$$

where $S_3^i$ and $S_3^j$ are bispectra of the $i$-th realization and the $j$-th realization. Here, $\Sigma$ in Eqs. (5a) through (5c) represents the summation over the frequencies $\omega_1$ and $\omega_2$ in the non-redundant region. Trivially, each type of the above defined BCCs yields 1 if we cross-correlate the same realization of the stochastic process with itself.

## 3. Demonstration by computer simulations

To demonstrate the effectiveness of the BCC method in stochastic processes, we consider *six* different types of *random telegraph signals* (RTS) and generate *seven* realizations of each type. The six different RTS types are as follows:

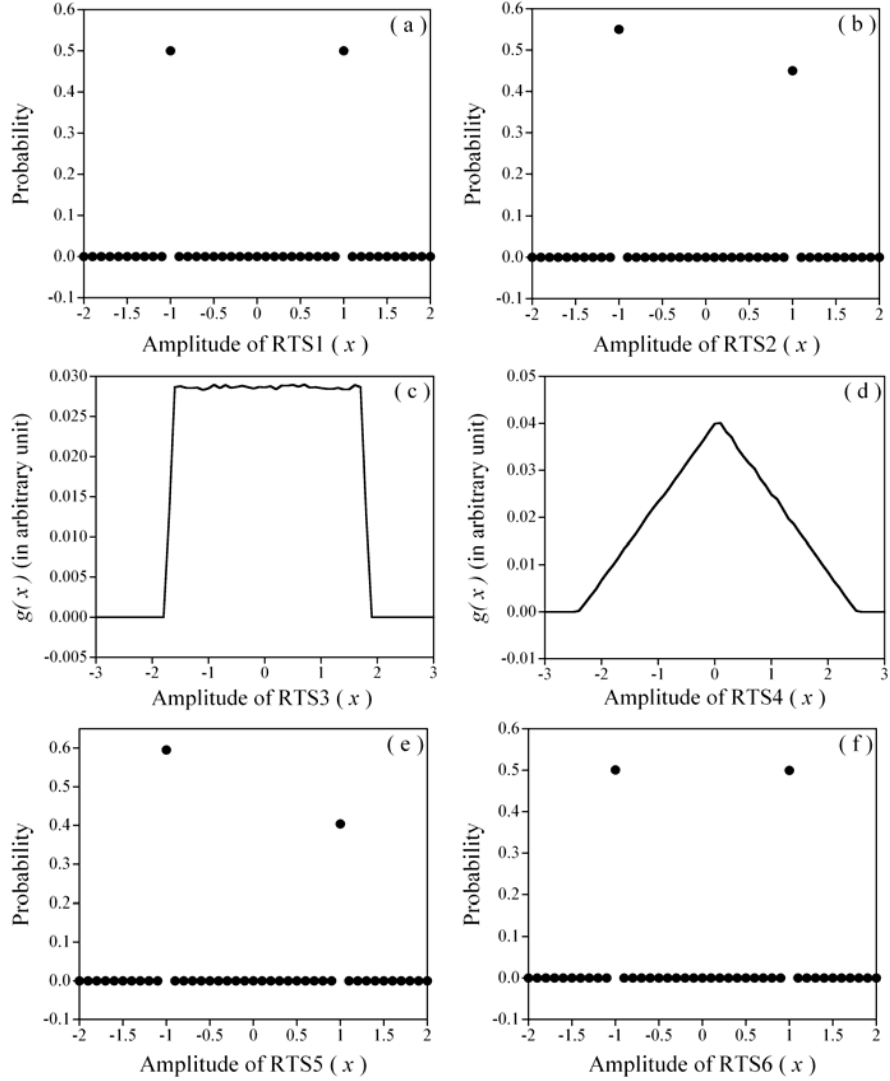

Fig. 1. Probabilities or the amplitude density functions $g(x)$ of the six random telegraph signals. The length of each sample is 1024 and the number of samples is 4096.

- RTS 1 has only two amplitudes, 1 and -1. The value is switched at Poisson time events with rate 0.2/step.
- RTS 2 has only two amplitudes, -1 and 1. The amplitude of the signal is changed by the combination of two Poisson rates depending on the amplitude of the sig-nal. The Poisson rate is 0.22/step at the amplitude 1 while it is 0.18/step at the amplitude -1.
- RTS 3 has continuous amplitudes randomly selected at each Poisson time events. The amplitude values are uniformly distributed in the range [-1.74, 1.74]. The Poisson rate is 0.4/step.
- RTS 4 has initial amplitude chosen randomly from the range [0, 2.45]. At a Poisson time event, the amplitude is changed as follows: if the amplitude is positive, the next amplitude is the previous one minus a random number uniformly distributed in the range [0, 2.45]. Otherwise, the next amplitude is the previous one plus a random number generated in the same way. Here, the Poisson rate is 0.4/step.
- RTS 5 has only two amplitudes, 1 and -1. The -1 amplitude is the ground state and the 1 amplitude belongs to a "*firing*" event. At a Poisson time event, firing is initiated. After a uniformly distributed random period, which is selected from the range of 1-11 steps, the firing is terminated and the system is reset to the ground state. Then a new Poisson timing starts for the initiation of the new firing event. The Poisson rate is 0.1/step.
- RTS 6 has only two amplitudes, 1 and -1. The signal holds its original amplitude for a random time period. After a random period uniformly distributed in the interval 1-26 steps, the sign is alternated.

The evaluation of the statistical properties of the RTS signals was obtained by computer simulations. The simulation length of each sample is 1024 and the number of samples is 4096. Figure 1 shows the probability or the amplitude density function (ADF) $g(x)$ of the six RTSs. Since the amplitude of RTS 1, RTS2, RTS 5, and RTS 6 is discrete, we need to use probability, not the amplitude density function. $P_1$ and $P_{-1}$ are probabilities for amplitude to be found 1 and -1 at a step, respectively. $P_1$ and $P_{-1}$ in RTS 1 is the same since the Poisson rate is constant. In RTS 2, $P_{-1}$ is bigger than $P_1$ since the Poisson rate depends on the amplitude. The amplitude density function of the RTS 3 is uniform since the amplitude is selected randomly in a uniform distribution and the Poisson rate is constant. The amplitude density function of the RTS 4 has a maximum at the amplitude zero and zero at the amplitude 2.45 or -2.45 since the smaller absolute value of the amplitude happens more frequently than the bigger. In RTS 5 case, since the condition for the amplitude 1 is different from that for the amplitude -1, $P_1$ and $P_{-1}$ is not the same. Since the random time period is independent of the amplitude, $P_1$ and $P_{-1}$ in RTS 6 is the same. These properties of the RTSs are shown in Figure 1 as expected. Figure 2 shows the PDSs of the six RTSs. It can be seen that RTS 1 through RTS 4 have the same PDS, but the PDSs of RTS 5 and RTS 6 are different.

Let us call auto-BCC the BCC between different realizations of the same type of stochastic process and cross-BCC the BCC between realizations of different stochastic processes, respectively. We generated 7 independent realizations for each RTS and evaluated the bispectrum for each signal. Thus the simulations provide 21 (=7*6/2) auto-BCCs for each RTS and 49 cross-BCCs for each pairs of RTSs.

We use a 3-dimensional plot (see Figure 3) so that the x-coordinate corresponds to the RBCC, the y-coordinate to the IBCC, and the z-coordinate to the MBCC, respectively. Thus the BCC plot of any realization of any stochastic process with itself would be a fixed point with coordinates (1,1,1). In Figure 3, the point groups labeled 'a' through 'f', represent the cross-BCCs while the points inside the ellipse represent the auto-BCCs.

The label 'a' represents the RTS 4-5 pair, the label 'b' the RTS 2-5 pair, the label 'c' the RTS 2-4 pair, the label 'd' the RTS 1-4, 1-5, 3-4, 3-5, 4-6, and 5-6 pairs, the label 'e' the RTS 1-2, 2-3 and 2-6 pairs, and the label 'f' the RTS 1-3, 1-6 and 3-6 pairs. The plots of almost all of the cross-BCCs, except that of the 2-4 pair, are around the z axis; that is, they have about zero RBCC and IBCC. The plots of the cross-BCCs of the 2-4 pair also have zero IBCC, but nonzero RBCC. Figure 4 shows the projection of the 3-dimensional BCC plot onto the RBCC-IBCC plane. The features of the BCC location are as follows:

- The plots of the cross-BCCs are around the z axis.
- The plots of the BCCs tend to form well distinguishable groups which tend to form characteristic shapes.
- The plots of the auto-BCCs tend to form lines and the cross-BCCs form 3-dimensional bundles. Because their locations and shapes are strongly different, they are clearly distinguishable from each other.

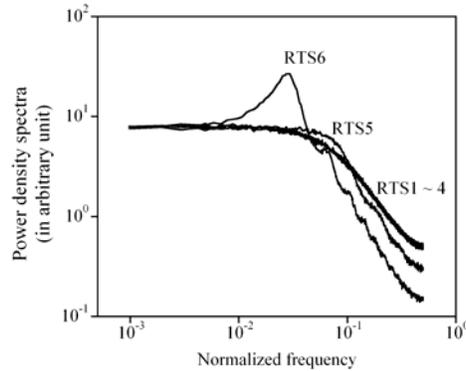

Fig. 2. Power density spectra of six random telegraph signals. They were obtained by direct conventional method without window.

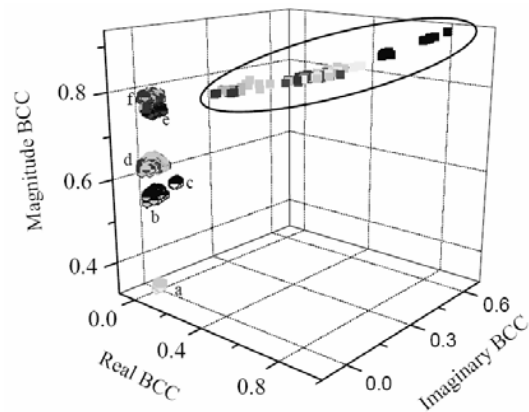

Fig. 3. Three-dimensional bispectrum correlation coefficient; the x-coordinate corresponds to the real-BCC, the y-coordinate to the imaginary-BCC, and the z-coordinate to the magnitude-BCC. The points inside the ellipse represent the auto-BCCs. The labeled *a* represents RTS 4-5 pair, the labeled *b* RTS 2-5 pair, the labeled *c* RTS 2-4 pair, the label d RTS 1-4, 1-5, 3-4, 3-5, 4-6, and 5-6 pairs, the labeled *e* RTS 1-2, 2-3 and 2-6 pairs, and the labeled *f* RTS 1-3, 1-6 and 3-6 pairs.

Therefore, the BCC method is a new type of pattern recognition tools. It is useful to determine which type of stochastic processes a realization belongs to. Conventional correlation technique is unable to do that. When we test if a certain realization x(t) of an unknown stochastic process is a realization of a known stochastic process K, we calculate and plot the BCCs of the x(t) process and several realizations of the process K. Then we get a bundle of points in the 3-dimensional space. The location and the shape of the bundle of the BCC points in Figure 3 and Figure 4 can be regarded as a classifier.

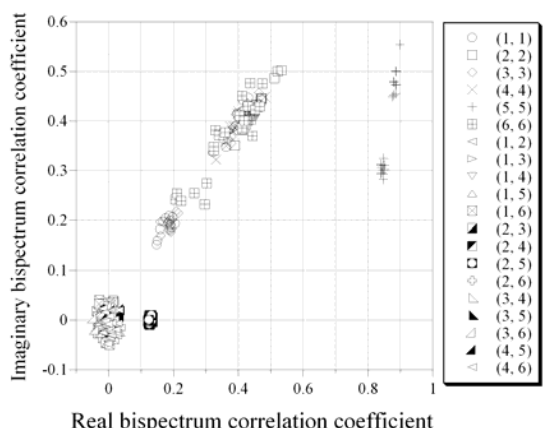

Fig. 4. Two-dimensional bispectrum correlation coefficient plot; the x-coordinate corresponds to the real-BCC, and the y-coordinate to the imaginary-BCC. The column at the right side shows the RTS pairs.

## 4. Summary

A new cross-correlation technique, which gives enhanced information about stochastic patterns, the bispectrum correlation coefficient method, was introduced. We demonstrated the usefulness of the BCC method with six types of random telegraph signals. This new pattern recognition has potential applications in biometrics, sensing, forensic, security, and image processing.

### Acknowledgement

J. U. Kim would like to acknowledge the support of Ebensbeger/Fouraker Graduate Fellowship.